\documentclass[
  ,draft            
  ]
  {aipproc}

\layoutstyle{6x9}

\begin{document}

\vspace*{-0.28in}
\begin{flushright}
LA-UR-09-05030
\end{flushright}
\vspace*{-0.31in}

\title{Epoch Dependent Dark Energy}


\author{B. H. J. McKellar}{
  address={School of Physics, University of Melbourne, Victoria 3010, Australia.}
}

\author{T. Goldman}{
  address={Theoretical Division, Los Alamos National Laboratory,\\
Los Alamos, New Mexico, 87545, USA}
}

\author{G. J. Stephenson, Jr.}{
  address={Dept. of Physics \& Astronomy, University of New Mexico,\\
Albuquerque, New Mexico 87131, USA}
 }
 
 \author{P. M. Alsing}{
  address={Dept. of Physics \& Astronomy, University of New Mexico,\\
Albuquerque, New Mexico 87131, USA}
 }

\begin{abstract}
 We present a model in which the parameter $w$ approaches $-1$ near a
particular value of $z$, and has significant negative values in a
restricted range of $z$.  For example, one can have $w \approx -1$ near $z = 1$,
and $w > -0.2$ from $z = 0$ to $z = 0.3$, and for $z > 9$.  The ingredients of the model are
neutral fermions (which may be neutrinos, neutralinos, etc) which are
very weakly coupled to a light scalar field.

This model emphasises the importance of the proposed studies of the
properties of dark energy into the region $z > 1$.
\end{abstract}

\maketitle
\section{Introduction}

About twenty five years ago the possibility of neutrinos interacting by exchange of very light neutral scalars to produce interesting cosmological and astrophysical effects was considered by Kawasaki, Murayama and Yanagida \cite{Kawasaki:1991gn}, and Malaney, Starkman and Tremaine\cite{Malaney:1994pn}.

Three of us later investigated a similar system, in which the scalar particle has a mass of order 1/a.u., which leads to neutrino clustering\cite{Stephenson:1996qj}.

Following the observation of the re-acceleration of the expansion of the
Universe~\cite{Riess:1998cb, Perlmutter:1998np}, which led to the concept of dark energy (for a recent review see ref.~\cite{Frieman:2008sn}).
Fardon, Nelson and Weiner~\cite{Fardon:2003eh} noted that the inferred energy density
of dark energy, of order  (2.4 meV)$^4$~\cite{Frieman:2008sn}, was, remarkably, of the
order of the experimentally inferred value of neutrino masses.  They also utilized the concept of a scalar field interacting with the neutrinos to obtain the desired effect. 

It is worth emphasizing that negative pressure is not a strange feature in physics, in that any self bound system has an equilibrium density, and will have a higher energy per particle as the density is decreased, and thus a negative pressure in this region.  Because this feature exists in our neutrino clustering model, we realized that it could produce negative values of $w$ and was thus  relevant to the dark energy problem. We provided some early comments \cite{McKellar:2004rs}, and then showed how it led to a system with $w \approx -1$  in a restricted range of the development of the universe~\cite{Goldman:2009wp}, with $w \to 1/3$ at very early times, and $w \to 0$ near the present.  In this paper we will be elaborating  our model of dark energy.  

To obtain the  energy density of the neutral fermion -- scalar field system as a function of $z$ one needs to make assumptions about parameters.  
Typically $w$ has a minimum as a function of $z$, and the position of the minimum may be moved by a choice of parameters, but a rapid variation of $w$ between the minimum and the present is characteristic of our model.  This lends support to attempts to probe the $z$ dependence of dark energy in more detail.

It is worth noting that  Mota et al\cite{Mota:2008nj, Pettorino:2009vn} have recently introduced the possibility of neutrino clustering in the neutrino -- scalar field model to obtain a dependence of the dark energy density on $z$.  Our work shows that a variation of $w$ and the dark energy density $\rho_E$  with epoch can be obtained in a homogeneous model.

After outlining the model, we display our results for $w$ as a function of the density of the neutral fermions, and then discuss how to choose the parameters of the model to obtain results for $w$ as a function of $z$, and discuss the implications of the results.

\section{The neutral fermion - scalar field model}

Following ref.~\cite{Stephenson:1996qj},
the equation of motion for a neutral fermion field, $\psi$, interacting with a scalar
field, $\phi$, is
\begin{eqnarray}
\left[\partial^2 + m_s^2\right]\phi & = & g\overline{\psi}\psi \label{PHI}\\
\left[i{/\!\!\!\!\!\partial} - m_n^{(0)}\right]\psi & = & -g \phi \psi,\label{PSI}
\end{eqnarray}
with  $\hbar = c = 1$.  The
nonlinear scalar selfcouplings are omitted here, even though
they are required to exist by field theoretic selfconsistency~\cite{McKellar:2004rs}, as they may
consistently be assumed to be sufficiently weak as to be
totally irrelevant.  
The parameter $m^{(0)}_n$ is the renormalized vacuum mass of the isolated neutral fermion, and takes into account the contributions from the electroweak theory, as well as contributions from the vacuum expectation value of the new scalar field $\phi$.

We look for solutions of these equations in infinite matter which are
static and translationally invariant. In the Freidman equation description of the evolution of the universe  we use the adiabatic approximation, assuming that the scalar field expectation value takes the value determined by the fermion density at the appropriate time\footnote{We have studied the corrections to the adiabatic approximation in the fermion-scalar field model and found them to be small, but have not yet investigated the corrections to this approximation when that model is coupled to the Freidman equation.}.  Equations (\ref{PHI}, \ref{PSI}) then lead to
an effective mass for
the
neutral fermion of
\begin{equation}
m^*_n = m^{(0)}_n - \frac{g^2}{m_s^2}\overline{\psi}\psi. \label{eq:effm}
\end{equation}
These equations are simply the equations of Quantum
Hadrodynamics
\cite{Serot:1984ey}.

As in \cite{Stephenson:1996qj},
we act with these operator equations on state which is a filled Fermi
sea of
neutral fermions, with a fermion number density $\rho_n$  and Fermi momentum
$k_F$,
related, for Majorana particles (which we assume henceforth), by $\rho_n~=~{k_{F}}^{3}/(3\pi^{2})$.  
We introduce the parameter
$
K_0~=~\left(g^2(m^{(0)}_n)^2\right)/\left(\pi^2m_s^2\right),
$
 and the variables
$y~=~m^*_n/m^{(0)}_n$, $ x~=~k/m^{(0)}_n$, 
$x_F~=~k_F/m^{(0)}_n$, and $e_F~=~\sqrt{x_F^2 + y^2}$.  Equation (\ref{eq:effm}) becomes an equation for $y$:
\begin{equation}
y = 1 - \frac{y K_0}{2}\left[e_Fx_F - y^2 \ln\left(\frac{e_F +
x_F}{y}\right) \right],\label{eq:YX}
\end{equation}

The total energy of the system,
$E = e {m^{(0)}_n}~ N$, with $N$ the number of fermions, and
\begin{equation}
e = \frac{3}{4} \left[ e_F + \frac{1}{K_0 x_F^3}(2-y)(1-y) \right]. \label{EPF}
\end{equation}
As $x_F$ goes from $0$ to $\infty$,  $y$ varies monotonically from $1$ to $0$, behaving as $y~\approx~1~-~\left(K_0 x_F^3\right)/3$ for small $x_F$, and $y \approx 2/\left(K_0 x_F^2\right)$ for large $x_F$.

For the fermion system to be bound, the minimum of
$e$
must be less than 1.
We find that for sufficiently large $K_0$ the neutral fermion -- scalar field system is bound~\cite{Stephenson:1996qj},
and the total energy displays the characteristic behaviour of a self-bound system noted above, suggesting that we should see negative pressures in this system.  We now go on to compute the equation of state parameter.

\section{Determining $w$}

The Equation of State, or the equation for the pressure, $P$, as a function of the other variables of the system, is represented through the equation of state parameter, $w$, which  is defined by the proportionality between $P$ and $\rho_E$, the energy density
\begin{equation}
P = w \rho_E. \label{EOS-1}
\end{equation}
As is well known, for radiation, $w = \frac{1}{3}$, for cold non-interacting
matter, $w = 0$, and, for a Cosmological constant, $w = -1$.  Any $w < 0$ describes
a system with negative pressure.

Using the equations of the Friedman-Lema\^{\i}tre-Robertson-Walker (FLRW)  metric, with a scale parameter $a$,
 the equation of state parameter is shown to satisfy
\begin{equation}
1 + w = - \frac{1}{3} \frac{\partial \ln \rho_E}{\partial \ln a}. \label{w1}
\end{equation}
In these equations $\rho_E$ is the \emph{total} energy density.  Assuming for this paper that the only source of this energy is the coupled fermion-scalar system, then
\begin{equation}
\rho_E =  e m_n^{(0)}  \rho_n, \quad \quad \rho_n \propto x_F^3 \quad \mbox{and} \quad x_F \propto a^{-1},
\end{equation}
and we obtain 
\begin{equation}
w = \frac{1}{3} \frac{\partial \ln( e )}{\partial \ln(x_F)} = \frac{1}{3} \frac{x_F}{e} \frac{\partial e}{\partial x_F}
\end{equation}
This is  the basic equation for computing $w$ in our model,
and it  gives
\begin{equation}
w = \frac{1}{3} \frac{e_F K_0 x_F^3 - 3(2 -y)(1 -y)}{e_F K_0 x_F^3   + (2 -y)(1 -y)} \label{eq:w}
\end{equation}
 It follows immediately that 
$w > -1$ in our model.

 \begin{figure}[h] 
\includegraphics[width=0.75\textwidth]{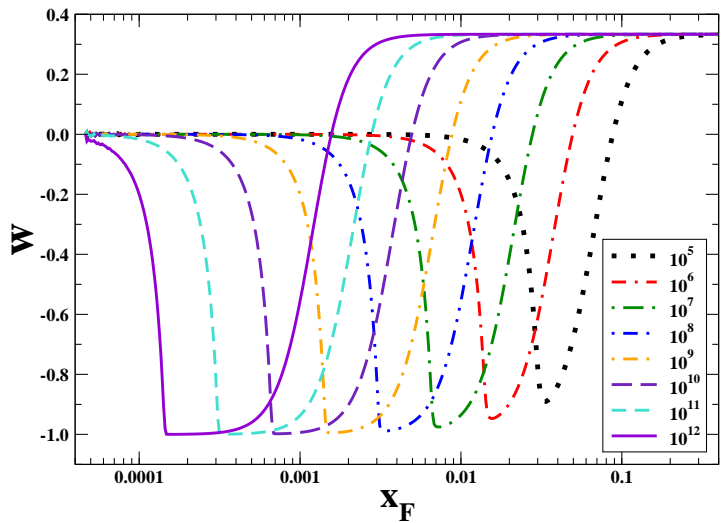}
\caption{$w$ vs. log($x_F$) for 8 values of $K_0$.}
\label{fig:wvslogxf}
\end{figure} 
 From equations (\ref{eq:YX},\ref{eq:w}), we can compute $w$ as a function of $x_F$, given $K_0$.  Figure \ref{fig:wvslogxf} shows some results using a log scale 
 for $x_F$.
 
Note that $w$ approaches     close to $-1$ as the density decreases (remember the direction of decreasing density is the direction of expansion of the universe) and then departs sharply towards zero,
At large $x_F$, it is clear that $w$ approaches $+1/3$ as it should for a relativistic
gas of fermions. The value
goes to zero at zero density. 

For large values of $K_0$ the minimum value of $w$ is very close to $-1$ and is reached near the value 
$x_F = x_{F,1} =  \left(3/K_0 \right)^{1/3},$
which can be obtained from an analytic approximation.  More precisely,  and empirically
$x_{F,m} \approx  \left( 3.82/K_0 \right)^{1/3}$.

\begin{table}[h]
\caption{
The minimum value of $w$, the point $x_{F,m}$ at which this is achieved, and the approximate location $x_{F,1}$
}
\label{tab:xfmink0} \vspace{6pt}
\begin{tabular}{|c|c|c|c|} 
\hline
log $K_0$ & log $x_{F,m}$   & log $x_{F,1}$& $w$ \\
\hline
6&-1.809&-1.841 & -0.94709\\
9&-2.806&-2.841& -0.99453\\
12&-3.806&-3.841 & -0.99945\\
\hline
\end{tabular}
\end{table}

The minimum value of $w$ approaches $-1$ more closely as $K_0$ increases, 
and the low density recovery of $w$ to zero becomes steeper as $K_0$ increases.

It is clear from these results that we have a model in which the value of $w$ is a function of the scale parameter $a$ or the red-shift $z$, i.e.  that we have an epoch dependent dark energy.  But these results are in terms of dimensionless parameters and variables.  To connect to the real world we must link these dimensionless numbers to the dimensionful numbers which characterize it.  As was shown in
ref. (\cite{Stephenson:1996qj}), there are only a few, weak constraints on the actual parameter
values. Moreover, very large values of $K_0$ are possible even for very small values of
$g^2$ if the range of the scalar is very large, corresponding to very small values
of $m_s$. Even if long-ranged, such weak interactions between fermions, especially
neutrinos or those outside of the Standard Model altogether (such as the LSP) are
exceptionally difficult to constrain by any laboratory experiments.

 \section{Connecting with dark energy, and fixing parameters}
 
 The energy density of dark energy is quoted as $\rho_{DE} = (3.20  \pm 0.4 ) \times 10^{-47}
\mbox{Gev}^4 $~\cite{Frieman:2008sn}  If we define $\rho_{DE} = m_E^4$, then $m_E = 2.4 $ meV.  It is important to remember that the conventional value of the dark energy density is derived on the assumption that it is constant during the evolution of the universe, and 
this is not the case in our model.  Pending a study of the development of the universe using our model we use the present estimate of $\rho_E$ and apply it to the region $w \approx - 1$ to estimate the relevant parameters in our model.
\begin{equation}
\rho_E  =  m_n^{(0)} e \rho_n =  \frac{(m_n^{(0)})^4 e x_F^3}{3 \pi^2} \label{e1} 
\end{equation}
To use this equation to estimate $m_n^{(0)}$ , first note that near $w \approx -1$, for large $K_0$, $x_F$ is small and we can use the appropriate approximations 
to get  
\begin{equation}
\rho_E \approx  \frac{(m_n^{(0)})^4  x_F^3}{6 \pi^2} \approx \frac{(m_n^{(0)})^4}{2\pi^2K_0},  \label{eq:re1}
\end{equation}
where, in the last equation, we use $x_F \approx  x_{F,1}$ at the minimum.  $K_0$ determines the relationship between $m_n^{(0)}$ and $m_E$, 
$m_n^{(0)} = \left[2 \pi^2 K_0\right]^{1/4} m_E$.
We choose $K_0$ , taking care that the implied values of $g^2$ and $m_s$ are reasonable.
For $K_0 = 10^6$, $m_n^{(0)} = 150 $~meV, in the neutrino range, and for $K_0 = 10^{54}$, $m_n^{(0)} = 150$~GeV, in the range expected for neutralinos.   What are the implications for the other parameters?  One can immediately estimate the density of the neutral fermions as  $47\times 10^3$~cm$^{-3}$ and $3.5 \times 10^{-8}$~cm$^{-3}$ for these two cases.

Because we are assuming that the neutral fermions form a homogeneous background, it is appropriate to set  the range of the scalar field to be the scale parameter of the universe at the relevant time, which we will take to be $z=1$,  \emph{ i.e.}
$m_s \sim \left(7 \times 10^9 {\rm\;  light years} \right)^{-1} \sim 3 \times 10^{-30} \; {\rm meV}.$
With this value of $m_s$ we can obtain the implied value of $g^2/(4\pi)$: $3\times 10^{-58}$ and $3 \times 10^{-54}$ in these two cases.
Even the largest coupling is far too weak to be constrained by terrestrial experiments.

An alternative way to proceed would be to assume a present density of the neutral fermions, $\rho_0$, and thus a present value of the Fermi momentum, $k_{F0}$, at a scale parameter $a_0$, use $a k_F = a_0 k_{F0}$ and $a_0 = (1+z)a$.  Equation (\ref{e1}) then gives
\begin{equation}
m_n^{(0)} = \frac{2}{(1+z)^3}\frac{\rho_E}{\rho_0} \label{fermimass}
\end{equation}
To give an explicit example, assume that the present density of neutral fermions is 100~cm$^{-3}$, characteristic of neutrinos, and extrapolate this to a density of 800 cm$^{-3}$ at $z=1$.  This gives a value $m_n^{(0)} = 9.2$eV, and $x_F = 3 \times 10^{-5}$, $K_0 = 9.7 \times 10^4$, and the even smaller coupling constant, $g^2/(4\pi) = 9 \times 10^{-64}$.

At this value of $K_0$, the minimum value of $w$ is $-0.8$, at the $1.5 \sigma$ level from the value $-1.05 \pm 0.18$ of the ESSENCE supernova survey\cite{WoodVasey:2007jb}, so even these parameters are not excluded.

It is impossible to make further progress without finding additional ways to constrain the parameters.  Given their extreme values the most promising approach is to use additional theoretical input, which we leave as a challenge for future work, by us and others.

\section{Results --- $z$ dependence of $w$}

Now that we have some parameters we can convert the determination of $w$ as a function of the dimensionless parameter $x_F$ to a dependence of $w$ on the red shift $z$.  To illustrate the results we have selected the values $10^6$ and $10^9$ for $K_0$, and set the minimum of $w$ to occur at $z = 1$.

\begin{figure}[h]
\includegraphics[width=0.75\textwidth]{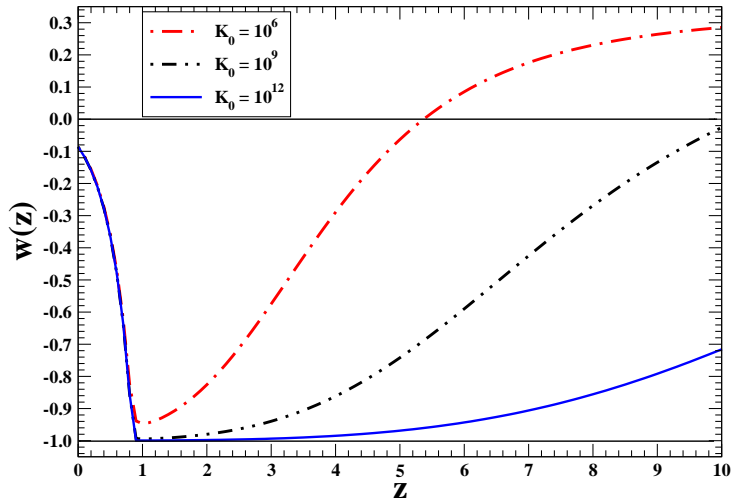}
\caption{$w$ vs. $z$ for  $K_0 = 10^6$, $K_0 = 10^8$ and $K_0 = 10^{12}$.}
\label{fig:wvz6}
\end{figure}

These results are illustrated in figure \ref{fig:wvz6}.

Note that there is a slow variation of $w$ with $z$ at epochs earlier than $z_{\mbox{{\small min}}}$ at which the minimum occurs, but a rapid change for $z$ between the present epoch ($z = 0$) and that of the minimum.

With the present uncertainty in data  we can choose to place the value of $z_{\mbox{{\small min}}}$ at any value of $z$ between, say $z = 0.5$ and $z = 1.2$ without being in conflict with the present data, and have chosen to show the results at $z_{\mbox{{\small min}}} = 1$ as illustrative results.

\section{Conclusions}

In an ideal future world one could imagine that $w(z)$ is indeed observed to have a minimum value at some $z_{\mbox{{\small min}}}$, and the the value of $w_{\mbox{{\small min}}}$, close to but greater than $-1$ is accurately known.  
The fact that the minimum value of $w$ is near $-1$ shows that  $K_0 > 10^{4.5}$, and the actual value of  $w_{\mbox{{\small min}}}$, if known accurately enough, would determine the value of $K_0$.
 Knowing the value of $K_0$ allows us to predict a value of $\rho_E$ at the minimum $w$, which provides  a good test on the model.  Then  a selection of parameters can place that minimum near an appropriate  $z$ value, $z_{\mbox{{\small min}}}$ .

Our prediction of a characteristic $z$ dependence of $w$ and $\rho_E$ leaves us vulnerable to developments in the precision of measurements and the extension of observations to larger and (with more difficultly) to smaller $z$, and we await with interest the results of the proposed experiments, such as those described in ref.~\cite{Frieman:2008sn}.

\section*{Acknowledgments}
This work was carried out in part under the auspices of the National Nuclear Security
Administration of the U.S. Department of Energy at Los Alamos National Laboratory
under Contract No. DE-AC52-06NA25396 and supported in part by the Australian
Research Council.

\end{document}